\documentclass[preprint]{aastex63}
\usepackage[normalem]{ulem}

\shorttitle{Neptune's obliquity}
\shortauthors{R. Gomes}

\usepackage{amsmath}

\begin{document}

\title{Neptune’s obliquity was likely engendered by Triton’s tidal evolution.}

\author{Rodney Gomes}
\affiliation{S\~ao Paulo State University, UNESP, Campus of Guaratinguet\'a, Av. Dr. Ariberto Pereira da Cunha, 333 - Pedregulho, Guaratinguet\'a - SP, 12516-410, Brazil, e-mail: rodney.gomes@unesp.br}

\begin{abstract}

	Neptune’s present axial tilt of approximately $28^\circ$ with respect to its orbital plane can be explained by collisions that its primordial core may have experienced with surrounding planetary embryos during the final stages of its formation. However, it is not certain that the ice giants underwent such a phase of intense collisions involving planetary embryos previously formed through the pebble accretion mechanism. Alternatively, Neptune could have attained its present mass solely through pebble accretion, without the formation of nearby planetary embryos capable of triggering catastrophic impacts. The embryo-collision scenario has the advantage of naturally explaining the large axial tilts observed in the ice giants. To account for these tilts without invoking late-stage catastrophic collisions, an alternative mechanism must be considered. In this work, I propose that if Neptune’s spin axis was initially nearly perpendicular to its orbital plane just after formation, its current axial tilt could result from the interaction between Triton’s tidally evolving orbit and Neptune’s spin axis, causing it to resonate with the solar system eigenfrequency $s_8$. Starting from a scenario in which a Triton-mass satellite is captured via the binary planetesimal disruption mechanism, I show that orbital evolutions bringing the satellite near Triton’s present orbit can induce a spin–$s_8$ resonance capable of producing a significant axial tilt of Neptune’s spin axis. To study this effect, I develop a model for planetary spin-axis evolution using Euler’s equations for a rigid body, which is incorporated into classical numerical integrations of the Newtonian equations of motion. I also include a tidal model to account for the satellite’s semimajor-axis decay and orbital circularization. Several numerical simulations are performed with this model, including Neptune as the central body, the newly captured satellite, the Sun and the three other giant planets. Increases in Neptune’s obliquity are observed, exceeding $50^\circ$ in some cases. An obliquity above $20^\circ$ is obtained for roughly $1/3$ of the cases. If Neptune initially had a near-zero obliquity, its current value could therefore have been naturally engendered by the tidal evolution of Triton.

\end{abstract}

\section{Introduction}

The formation of planetary systems from the gravitational collapse of a giant molecular cloud naturally implies a preferential direction of rotation for all objects within the forming planetary system, as angular momentum is conserved during the collapse. This preferential rotation direction encompasses not only the planets orbiting the central star, but also the satellites orbiting their planets, as well as the rotational axes of both planets and satellites. Indeed, this pattern is observed in the Solar System, albeit with important exceptions. The most notable examples are Triton's retrograde orbit, Venus's upside-down spin axis, and Uranus's sideways rotation. Neptune’s rotation axis, tilted by about $28^\circ$ relative to the Solar System's invariable plane, is less extreme than Uranus's, but still requires an explanation beyond the general trend set by the collapse of the primordial molecular cloud.

Explanations for the exceptional rotational directions of Solar System bodies must be rooted in the processes that shaped the early Solar System, particularly planetary accretion from smaller components such as planetary embryos, planetesimals, and pebbles. Planetary formation theories predict that once sufficiently large planetesimals form, they can accrete inward-drifting pebbles through gas drag to build planetary cores \citep{lambrechtsandjohansen2012, morbidelliandnesvorny2012, kretkeandlevison2014}. The resulting spin axis of a planet formed primarily through pebble accretion is expected to remain close to its primordial orientation—nearly perpendicular to the Solar System's invariable plane \citep{donesandtremaine1993, johansenandlacerda2010}.

To explain the conspicuous spin axis inclinations of the ice giants, \citet{izidoroetal2015} proposed that while pebble accretion was efficient in forming planetary embryos smaller than Uranus's and Neptune's cores, subsequent collisions between these embryos eventually produced the ice giants' final cores. There is ongoing debate about whether the formation of the ice giants involved collisions between embryos or relied solely on pebble accretion \citep{morbidelli2020, izidoroetal2021, savvidouandbitsh2023}. However, if we adopt the model in which the ice giants' cores formed solely through pebble accretion, we must invoke an external mechanism to explain the significant tilts of their spin axes.

Uranus's spin axis was proposed to have been tilted by the perturbation of a putative large primordial satellite \citep{boueandlaskar2010}. However, a major difficulty with this hypothesis was explaining the subsequent disappearance of this satellite while preserving (or forming) Uranus’s current satellite system. More recently, \cite{saillenfestetal2022} proposed instead that a former satellite of Uranus evolving through one giga-year-long tidal migration could have produced Uranus’s present tilt, while the satellite became destabilized when Uranus reached an obliquity above $80^\circ$ and eventually collided with the planet.
\cite{Millholland-Batygin2019} proposed a scenario based on the nodal precession that planets experience due to the gravitational potential of the protoplanetary disk in which they formed. However, this method can hardly explain the obliquities of the ice giants since large-scale obliquity excitation was likely inhibited by gravitational planet–planet perturbations. \cite{rogoszinski-hamilton2020} proposed that a sufficiently massive protoplanetary disk could induce nodal precession of the ice giants that would resonate with their spin-axis precession. This mechanism could increase Uranus’s obliquity to $70^\circ$, and a further tilt to its present value could have been attained through collisions with small planetary embryos. Neptune’s obliquity could reach $30^\circ$ with a more realistic original circumplanetary disk.
The obliquity of Jupiter has been proposed to have been produced by spin–orbit resonances induced by the tidal migration of the Galilean satellites \citep{saillenfestetal2020, dbouk-wisdom2023}, whereas Saturn’s obliquity has been proposed to have been induced by its current regular satellites \citep{saillenfestetal2021a, saillenfestetal2021b, saillenfest-lari2021, wisdometal2022, cuketal2024}. In all these cases, the tilting process of Saturn was proposed to be due to a capture in the spin-$s_8$ resonance due to the tidal migration of Titan. The spin-$s_8$ resonance has also been claimed to have excited Saturn’s obliquity during the late stages of Neptune’s past migration \citep{ward-hamilton2004, hamilton-ward2004}. An analytical development of how this resonance operates can be found in \citep{ward-hamilton2004}.

Here I propose a scenario to tilt Neptune’s spin axis, assuming its initial obliquity was very low, without adding any additional ad hoc assumptions, and based solely on the tidal evolution of its existing main satellite, Triton. The proposed scenario must be associated with a particular model for the capture of Triton as Neptune's satellite. Theories of satellite formation during gas accretion onto giant planets generally predict prograde orbits. Therefore, the capture of satellites from originally unbound orbits has been the most frequently explored mechanism to explain satellites in retrograde orbits \citep{mckinnon1984, goldreich1989, turrinietal2008, turrinietal2009}.

However, capture is an unlikely event, as it requires either an efficient energy dissipation mechanism or three-body interactions to transfer energy from the captured body to another object. \citet{nesvornyetal-2007} proposed that irregular satellites (namely, satellites of giant planets with large orbital inclinations relative to the equatorial plane) were captured during planet-planet encounters, which occurred during the phase of giant-planet orbital instability \citep{tsiganis-a-2005} within a dispersed planetesimal disk. While this capture scenario can reproduce most of the known orbits of irregular satellites around the major planets, it is unlikely to explain the capture of a large object like Triton, given the small capture probability (on the order of $5 \times 10^{-7}$) and the limited expected number of Triton-sized objects in the planetesimal disk (at most a few thousand). \citet{Agnor-Hamilton2006} proposed a radically different scenario, based on the capture of one member of a binary planetesimal during the era when the giant planets were still embedded in the planetesimal disk. This scenario is viable, and even likely, provided that binary objects with components as massive as Triton or Pluto existed. More recently, \citet{gomesandmorbidelli2024} proposed that Triton was originally a regular satellite of Neptune. Through several collisions between proto-Neptune and planetary embryos, Neptune’s spin axis could have been tilted to a large inclination and then relaxed back to its current $30^\circ$, inducing a significant inclination of Triton's orbital plane relative to Neptune’s equator.

Here, I adopt the binary planetesimal capture scenario to explain Triton's orbit. Although \citet{gomesandmorbidelli2024} provides a feasible mechanism to produce Triton’s orbit, the binary planetesimal capture mechanism, differently from \citet{gomesandmorbidelli2024}, produces the initially high eccentricity and large semimajor axis orbits for Triton that are necessary to make the mechanism presented in this paper possible. This work thus consists of a series of simulations, where Triton is introduced as a captured object at some point during planetary migration within a planetesimal disk. In this framework, Triton initially occupies an inclined and highly eccentric orbit, evolving by tidal interactions to its present orbit around Neptune. The simulations incorporate perturbations from Neptune’s non-spherical shape, the Sun, and the other major planets. Several models have been so far developed to study the variation of spin-axes (see, for instance, \cite{vaillantetal2019} that gives an instructive summary of available averaged and non-averaged methods to study spin-axis variation). Anyway I decided to develop my own method to model the evolution of Neptune’s spin axis during the system's evolution, by integrating Euler’s equations for a rigid body together with the Newtonian equations of motion. This model thus considers the non-averaged equations for the variation of the planet's spin axis, which provide the necessary precision when dealing with orbits of short orbital period like those of satellites that influence the planet's spin axis. This model is presented in the Appendix, where it passed through a rigorous validation procedure. In Section 2, I describe simulations involving Neptune as the central body, Triton in a retrograde orbit, and the other three major planets, without tidal forces. I then identify particular Triton orbits that, along with the gravitational influence of the other bodies, induce a secular spin-$s_8$ resonance. In Section 3, using the results from Section 2 that identify Triton orbits capable of inducing the spin-$s_8$ resonance, I introduce tides into the simulations to search for cases where Triton passes through the resonance-inducing region. This is done progressively over three subsections: first considering only Neptune and Triton, then including the Sun, and finally including the other major planets. Section 4 presents the conclusions.

\section{Variation of Neptune's obliquity}

\begin{figure}
\centering
\includegraphics[scale=.30, angle=-90]{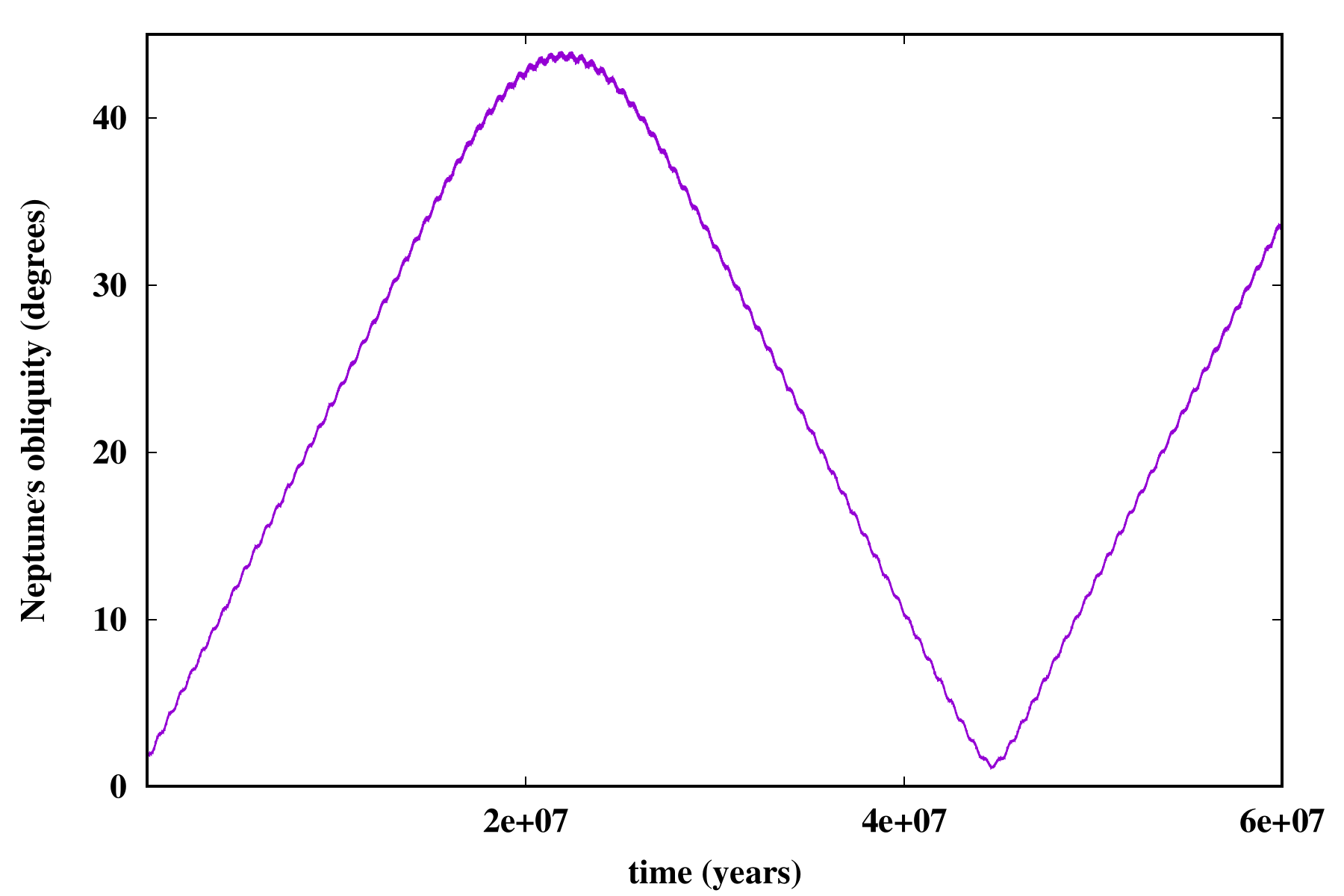}
        \caption{Variation of the obliquity of Neptune.}
        \label{fig-oblnet-semmare}
\end{figure}

\begin{figure}
\centering
\includegraphics[scale=.30, angle=-90]{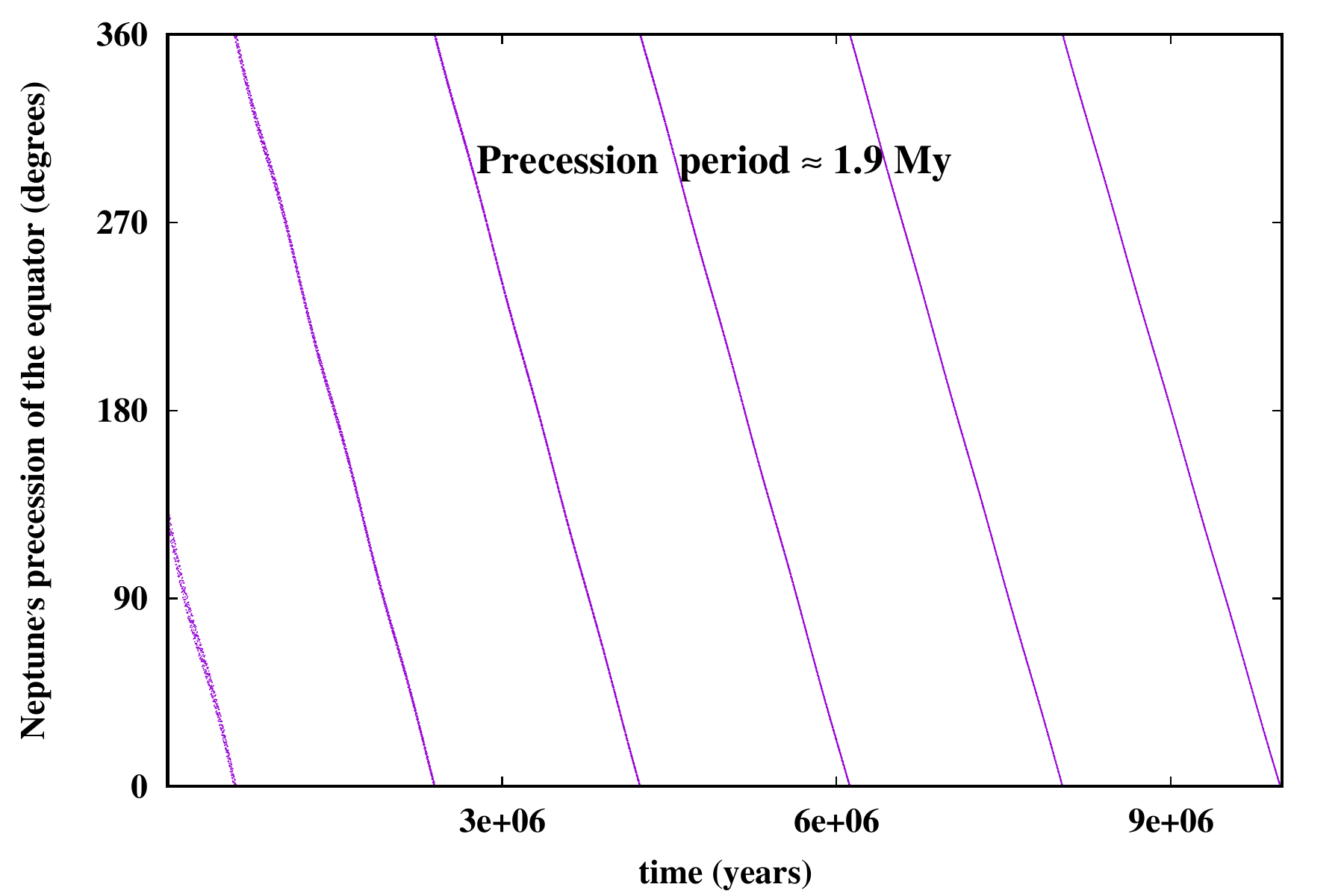}
        \caption{Variation of the precession of Neptune's equator.}
        \label{fig-precnet-semmare}
\end{figure}
\subsection{Parameters and initial conditions}

I perform several sets of simulations using the previously validated integrator, always considering Neptune as the central body and a Triton-mass satellite. In addition to Neptune and the satellite, some sets of simulations include the Sun, while others include the Sun and the three other major planets. The Triton-mass satellite is initialized in an eccentric, retrograde orbit based on the binary planetesimal capture model \citep{Agnor-Hamilton2006}. Neptune is assumed to start with zero obliquity when no planets are included, or with a small obliquity in some cases where the other major planets are included\footnote{In this case, Neptune’s equator is initially aligned with the reference plane, following the method described above. The initial reference plane is taken to be the ecliptic J2000; thus, Neptune’s obliquity starts at approximately $1.7^\circ$ with respect to Neptune’s orbital plane. Although the ecliptic has no physical role in these integrations, this choice provides a convenient way to initialize Neptune with a small, nonzero obliquity based on current planetary orbital elements.}. The positions of the Sun and planets relative to Neptune are taken from the JPL Horizons system. Other specific details for each set of simulations are provided in the sections where these simulations are discussed.

Although these simulations are intended to represent an early phase of Solar System evolution—during planetary migration within a planetesimal disk—the exact timing of Triton’s capture is uncertain. Moreover, adding a final phase of planetary migration during Triton’s evolution might unnecessarily complicate the simulations. I therefore decided to use the relative coordinates and velocities of the planets referenced to a recent date. Additional parameters are listed in Table \ref{parameters}.

\begin{table*}
	\begin{center}
        \caption{Planets' and satellite's parameters.}
\label{parameters}
\resizebox{\textwidth}{!}{%
		\begin{tabular}{lcccc}
\\
\tableline\tableline
	&  & Masses (in Earth masses)&  &  \\
\tableline
Satellite   &  Neptune &  Jupiter &  Saturn &  Uranus  \\
			$0.356252\times10^{-2}$ & $17.015$ & $318.098$ & $95.2591$ & $14.222$ \\
\tableline
	& & Neptune's shape and spin parameters && \\
\tableline  
	$J_2$  & & 0.0034  &  &    \\
	Equatorial radius & & 24764.8 km &   &  \\
	Spin frequency  & & 3442.4 $yr^{-1}$  &  &  \\
	Moment of inertia factor &  & 0.23 &   &   \\
\tableline
        & & Other satellite's and Neptune's parameters for tidal evolution. && \\
\tableline
	$R_S$ & &  $1352.98$ km       & & \\
	$\omega_s$  & &  $3000$ yr$^{-1}$         & & \\
	$\epsilon_s$  & & $0$        & & \\
	${K_2}_s$     & & $0.1$        & & \\
	${\Delta t}_s$  & & $808$ s      & & \\
        ${K_2}_p$     & &   $0.407$      & & \\
        ${\Delta t}_p$  & & $1.02$ s      & & \\
\tableline

\end{tabular}
		}
\end{center}
\end{table*}

\begin{figure}
\centering
\includegraphics[scale=.30, angle=-90]{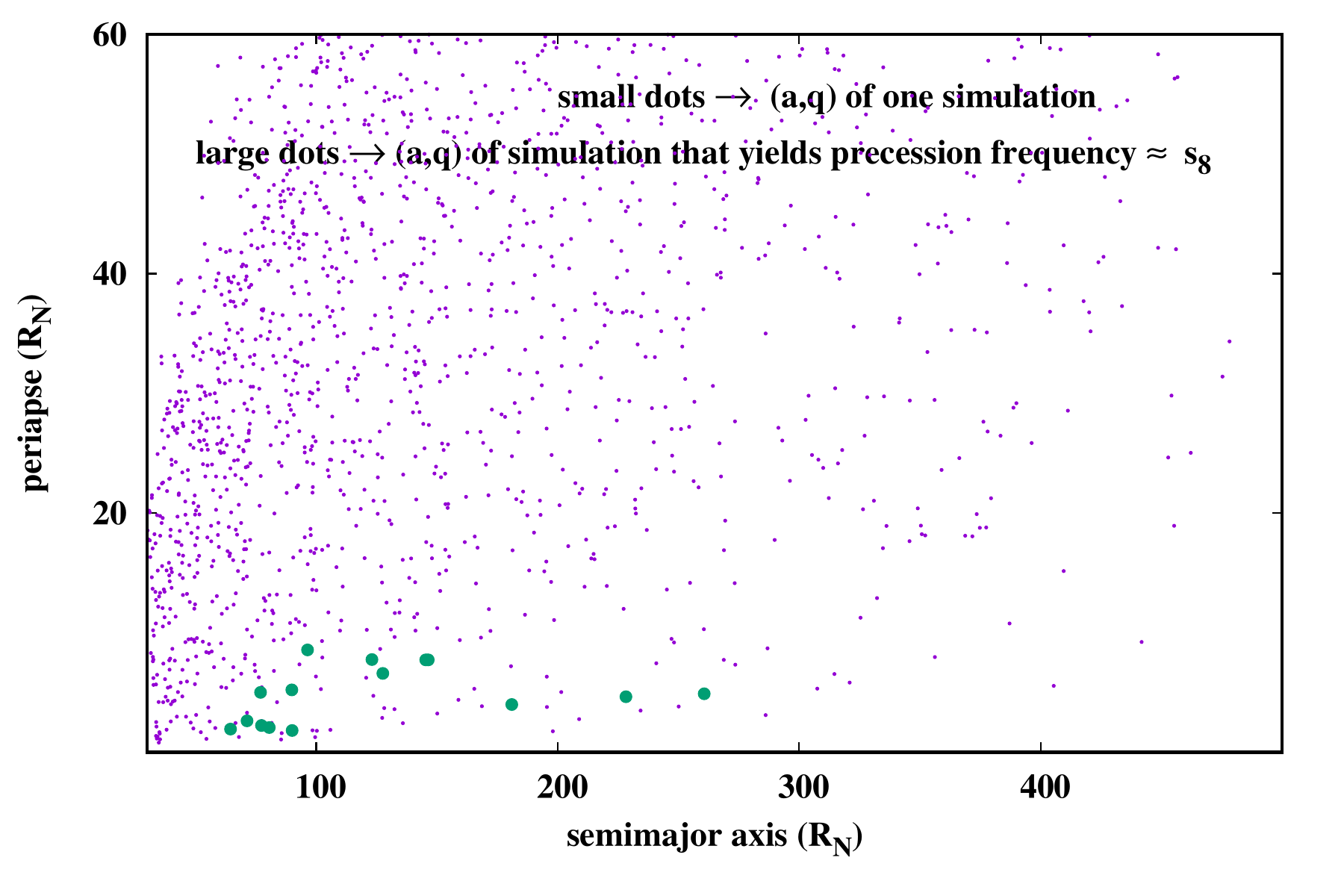}
        \caption{Mean semimajor axes and periapses of satellites around Neptune simulated in a numerical integration with the Sun and the other three major planets. Large dots correspond to numerical integrations that yilded Neptune's precession frequency commensurable with the Solar System proper frequency $s_8$.}
        \label{fig-ressecs8-amplo}
\end{figure}

\begin{figure}
\centering
\includegraphics[scale=.30, angle=-90]{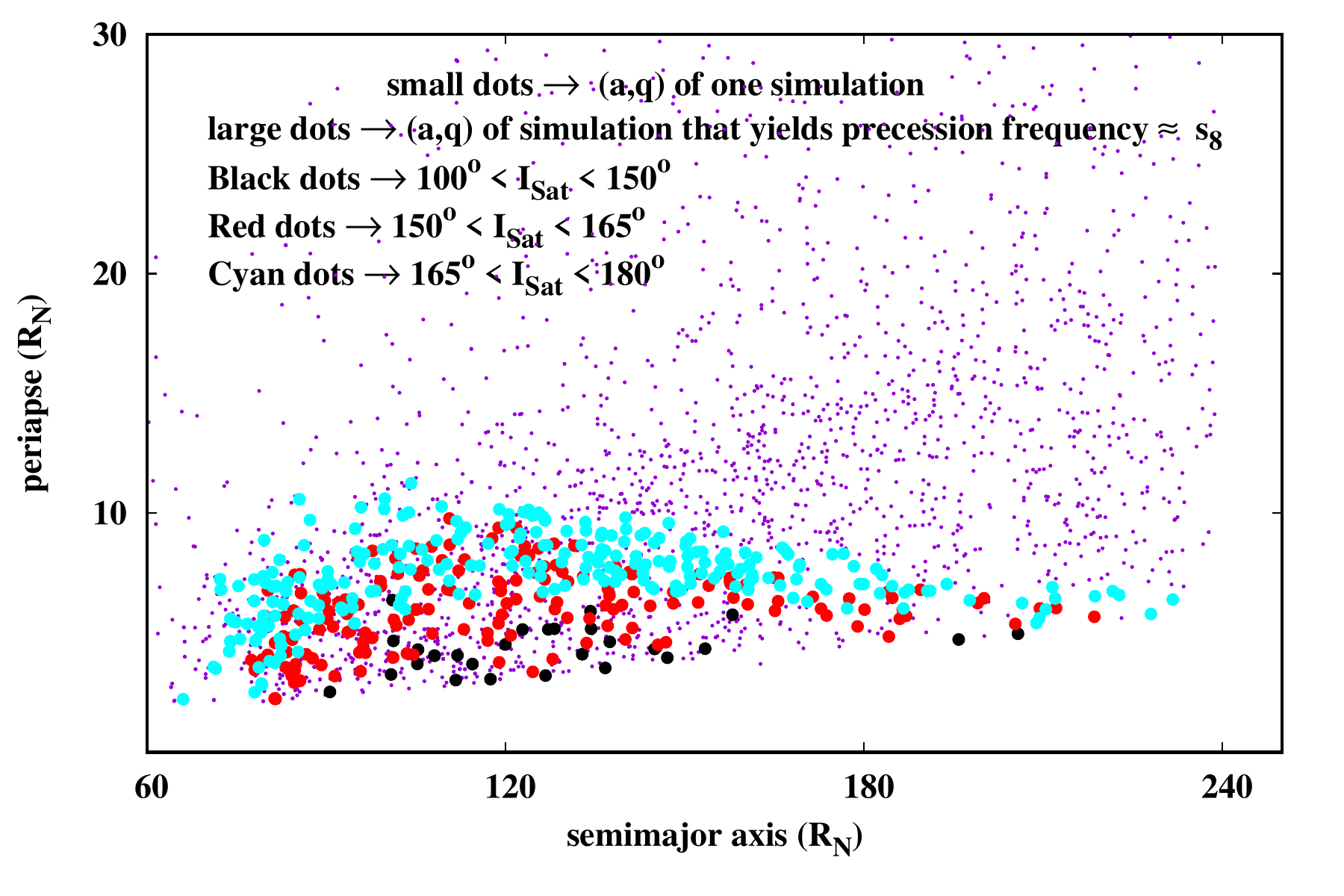}
	\caption{Same as Fig. \ref{fig-ressecs8-amplo} for simulations in a smaller range of semimajor axes and periapses.}
        \label{fig-ressecs8-restr}
\end{figure}

\subsection{Seeking secular resonances with $s_8$}

I begin a first set of simulations considering Neptune, the Sun, and the other three major planets. These initial simulations are intended to explore the behavior of Neptune's obliquity for various Triton-like orbits. The satellite’s retrograde orbit is initialized with semimajor axes ranging from $60$ $R_N$ to $300$ $R_N$, eccentricities between $0.9$ and $0.98$, and inclinations from $100^\circ$ to $180^\circ$. By examining the evolution of Neptune’s obliquity in these simulations, I found some interesting results, such as the one shown in Fig. \ref{fig-oblnet-semmare}, which illustrates Neptune’s obliquity evolution. Investigating the cause of Neptune’s obliquity increasing to above $40^\circ$, I found that the precession period of Neptune’s equator for the same case was approximately $1.9\times 10^6$ years as shown in Fig. \ref{fig-precnet-semmare}— close to the Solar System’s eigenfrequency $s_8$. These cases of increased obliquity motivated a new series of simulations, using similar initial orbital elements for Triton as in the exploratory integrations above, but now with a somewhat broader range. The new simulations consider initial semimajor axes for Triton ranging from $30 R_N$ to $500 R_N$, and eccentricities from $0.2$ to $0.99$, while initial inclinations remain in the same range as before. I also restrict the total integration time to $5\times10^6$ years, which is sufficient to detect a precession frequency of Neptune’s equator near $s_8$. The outcomes are analyzed using frequency analysis to identify cases in which Neptune’s precession period lies within the range of $1.7 \times 10^6$ to $2.1 \times 10^6$ years. Figure \ref{fig-ressecs8-amplo} plots the mean semimajor axis and mean periapsis of the satellite for each simulation, with larger dots representing cases where the precession period falls within the specified range.

\begin{figure}
\centering
\includegraphics[scale=.30, angle=-90]{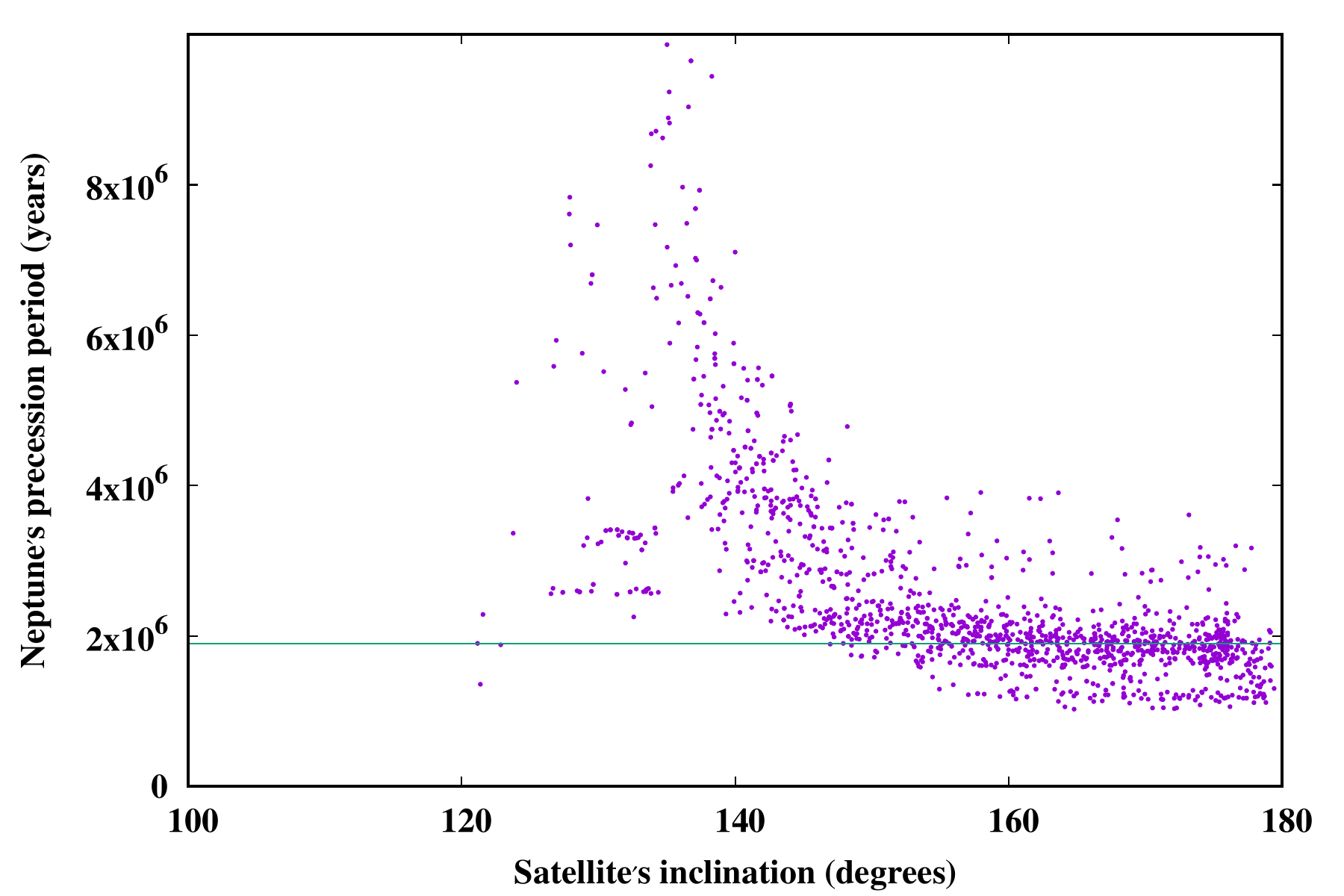}
        \caption{Relation between the satellite's inclination with respect to Neptune's equator and the precession period of Neptune's equator for all satellites with semimajor axes in the range $60$ and $240 R_N$ and periapses smaller than 10 $R_N$. The horizontal line indicates the period associated with $s_8$ solar system proper frequency.}
        \label{fig-Iper-restr}
\end{figure}

These results prompted further simulations with narrower ranges of semimajor axes and eccentricities, corresponding to the regions populated by large dots in Fig. \ref{fig-ressecs8-amplo}. The new ranges are $60 R_N$ to $240 R_N$ for the semimajor axes and $0.9$ to $0.98$ for the eccentricities. Figure \ref{fig-ressecs8-restr}, similar to Fig. \ref{fig-ressecs8-amplo}, shows the results with more detail in the region that produces resonances between Neptune’s precession and the Solar System eigenfrequency. This figure also highlights the influence of the satellite’s inclination relative to Neptune’s equator—larger inclinations result in a higher number of simulations where Neptune’s precession frequency approaches $s_8$. However, not all satellites in the region marked by large dots induce a precession frequency near $s_8$; this also depends on other orbital elements such as the satellite’s longitude of ascending node and argument of periapsis. Nevertheless, for all satellites in the region defined by $60$ au $< a < 240$ au and $q<10$ au, $33.6$\% induce a precession frequency in the range $1.7\times 10^6$ to $2.1\times 10^6$ years. Expanding this range by $0.1 \times 10^6$ years on both sides increases this proportion to $47.0\%$. If we restrict the sample further to satellites with $I>150^\circ$, these proportions increase to 48.4\% and 66.3\%, respectively. Figure \ref{fig-Iper-restr} shows the relationship between the satellite's inclination relative to Neptune’s equator and the precession period of Neptune’s equator for all satellites with semimajor axes in the range $60$ and $240$ $R_N$, and periapses less than $10$ $R_N$. This confirms that resonances between Neptune’s precession frequency and the Solar System eigenfrequency $s_8$ are primarily induced by satellites with high inclinations, semimajor axes between $60$ and $240$ $R_N$, and periapses smaller than $10$ $R_N$.

\subsection{Introduction of tides: Linking Tritons's evolution with the $s_8$ secular resonance}

The results from the previous section naturally raise the question of whether satellites captured by Neptune through the binary planetesimal capture model would pass through the resonance-inducing region shown in Fig. \ref{fig-ressecs8-restr}. This motivates a series of simulations incorporating a model for the satellite’s tidal evolution. I adopt a simple tidal model to study the evolution of the satellite’s semimajor axis and eccentricity, following the approach by \cite{Mignard1979, Mignard1980} and \cite{Hut1982}, the same procedure used by \cite{correia2009} and \cite{nogueiraetal2011}. We integrate the equations governing the evolution of the satellite’s semimajor axis $a$, eccentricity $e$, inclination with respect to Neptune's equator $I$, spin rate $\omega_s$, and obliquity $\varepsilon_s$, while neglecting the tidal influence on Neptune’s rotation rate and obliquity. The equations used are:

$$
\begin{aligned}
	\dot{a}= & \frac{2 K_{s}}{m_{s} a}\left(\frac{\omega_{s}}{n} F_{2}\left(e\right) \cos{\varepsilon_{s}} - F_{3}\left(e\right)\right) \nonumber \\
	& +\frac{2 K_{p}}{m_{s} a}\left(\frac{\omega_{p}} {n} F_{2}\left(e\right) \cos{I} -F_{3}\left(e\right)\right), \nonumber \\
	\dot{e}= & \frac{9 K_{s} e}{m_{s} a^{2}}\left(\frac{\omega_s}{n} \frac{11 F_{4}\left(e\right) \cos {\varepsilon_{s}} }{18}-F_{5}\left(e\right)\right) \nonumber \\
	& +\frac{9 K_{p} e}{m_{s} a^{2}}\left(\frac{\omega_p}{n} \frac{11 F_{4}\left(e\right) \cos {I}}{18}-F_{5}\left(e\right)\right), \nonumber \\
        \dot{\omega}_{s}= & -\frac{K_{s} n}{C_{s}}\left(\frac{\omega_{s}}{n} F_{1}\left(e\right) \frac{1+\cos ^{2} \varepsilon_{s}}{2} -F_{2}\left(e\right) \cos \varepsilon_{s}\right), \nonumber \\
	\dot{\varepsilon}_{s}= & \frac{K_{s} n}{C_{s} \omega_{s}} \sin \varepsilon_{s}\left(\frac{\omega_{s}}{n} \frac{F_{1}\left(e\right) \cos \varepsilon_{s}}{2}-F_{2}\left(e\right)\right), \nonumber 
\end{aligned}
$$

\noindent where,

$$
\begin{aligned}
K_{s}= & \frac{3 k_{2 s} G m_{p}^{2} R_{s}^{5} \Delta t_{s}}{a^{6}}, \nonumber \\
K_{p}= & \frac{3 k_{2 p} G m_{s}^{2} R_{p}^{5} \Delta t_{p}}{a^{6}}, \nonumber \\
F_{1}(e)= & \left(1+3 e^{2}+3 e^{4} / 8\right)\left(1-e^{2}\right)^{-9 / 2}, \nonumber \\
F_{2}(e)= & \left(1+15 e^{2} / 2+45 e^{4} / 8+5 e^{6} / 16\right)\left(1-e^{2}\right)^{-6}, \nonumber \\
F_{3}(e)= & \left(1+31 e^{2} / 2+255 e^{4} / 8+185 e^{6} / 16+25 e^{8} / 64\right) \left(1-e^{2}\right)^{-15 / 2}, \nonumber \\
F_{4}(e)= & \left(1+3 e^{2} / 2+e^{4} / 8\right)\left(1-e^{2}\right)^{-5}, \nonumber \\
F_{5}(e)= & \left(1+15 e^{2} / 4+15 e^{4} / 8+5 e^{6} / 64\right)\left(1-e^{2}\right)^{-13 / 2} . \nonumber 
\end{aligned}
$$
\vspace{1.5cm}

In these equations, the subscript $s$ refers to the satellite and $p$ to the planet, $n$ is the mean motion, $\omega$ the spin rate, $\varepsilon$ the obliquity, $k_2$ the Love number, and $\Delta t$ the tidal response time. The satellite’s moment of inertia along its spin axis is given by $C_s = 0.35 m_s R_s^2$, where $R$ is the equatorial radius and $m$ the mass. The constants used in these formulae are the same as those in \citep{nogueiraetal2011} and are given in Table \ref{parameters}.

\subsubsection{Just the satellite and Neptune}

\begin{figure}
\centering
\includegraphics[scale=.30, angle=-90]{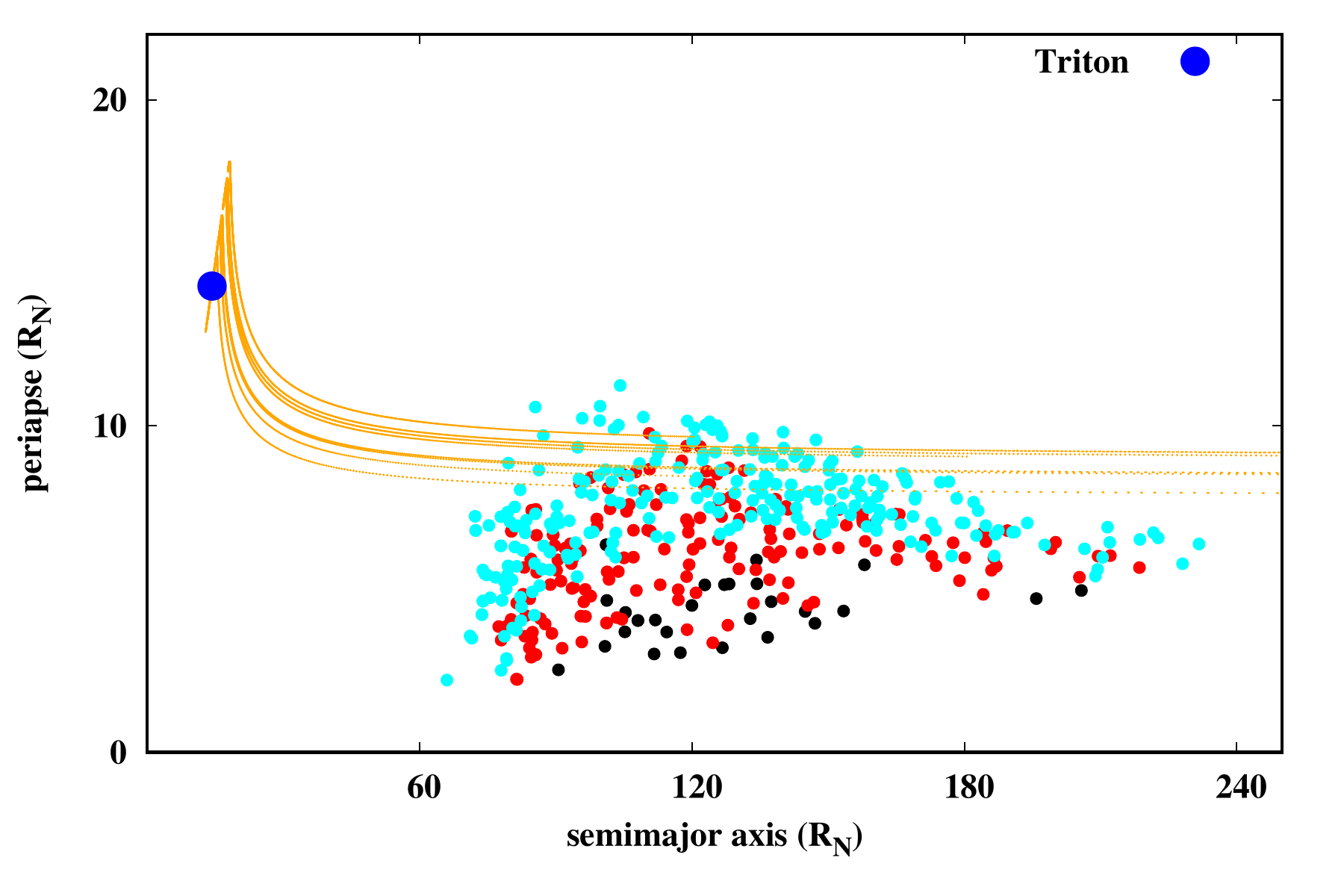}
	\caption{Evolution of the semimajor axis and periapsis of several satellites affected by tides (orange dots) plotted over the map showing the places where resonances of Neptune's precession frequency with $s_8$ take place as depicted in Fig. \ref{fig-ressecs8-restr}. The large blue circle stands for Triton}
        \label{fig-somare-restr}
\end{figure}

The first set of simulations with tides includes only the planet and the satellite. Tidal evolution is governed by the equations above, as if only the satellite and Neptune were present. I begin the integrations with the satellite’s semimajor axis varying from $0.02$ au to $0.25$ au in steps of $0.01$ au (approximately $120 R_N$ to $1500$ $R_N$, with a step of $60$ $R_N$), eccentricities ranging from $0.8$ to $0.99$ in steps of $0.01$, and inclinations from $100^\circ$ to $180^\circ$ in steps of $10^\circ$ \footnote{The satellite's inclination does not vary in this simplified model, but it influences the evolution of the semimajor axis and eccentricity.}. These ranges roughly correspond to those of a satellite just captured through the binary planetesimal model \citep{nogueiraetal2011}. I integrate the equations for $4.5$ Gyr and examine the final semimajor axes and eccentricities of the satellites. Simulations resulting in a semimajor axis below $0.003$ au ($\sim 18 R_N$), eccentricity below $0.001$, and inclination greater than $150^\circ$ are considered Triton-generating cases. Figure \ref{fig-somare-restr} shows the evolution of the semimajor axes and periapses for these Triton-generating cases, plotted over the large dots that represent the region of $s_8$-inducing resonance as shown in Fig. \ref{fig-ressecs8-restr}. We observe that in all such cases, the satellite crosses the resonance-inducing region. I repeated the above simulations using $\Delta t_s = 8080\, s$, which better corresponds to a semi-molten Triton during its tidal circularization phase \citep{mckinnon1984}. The results are qualitatively similar, as all satellites satisfying the constraints mentioned above still cross the resonance-inducing region. The only difference is that, with $\Delta t_s = 8080\, s$, fewer satellites meet the final semimajor axis and eccentricity criteria.

\subsubsection{Satellite, Neptune and the Sun}
I now introduce the perturbations from the Sun to examine the satellite’s evolution under tidal effects and to determine whether the satellite will pass through the region where the secular spin–$s_8$  resonance is active. In this simplified model, the $s_8$  proper frequency does not appear, but the satellite’s evolution is not expected to differ significantly from that predicted by the more complete model, which includes the other three major planets. The most notable consequence of the Sun’s perturbation is the appearance of the Kozai resonance, which can induce significant coupled variations in the satellite’s eccentricity and inclination. An immediate consequence of eccentricity fluctuations—particularly for a satellite whose orbit already starts with high eccentricity, as predicted by the binary planetesimal capture model—is the possibility of an imminent collision with Neptune. Furthermore, the averaged tidal model presented earlier may not be accurate when the Sun is included in the system. The Sun can induce a significant and rapid increase in the satellite's eccentricity while it is near apoapsis, leading to a fast semimajor axis decay predicted by the averaged tidal equations. However, this is incorrect, as the actual decay occurs primarily at periapsis. In such cases, the averaged formalism used to compute tidal variations of the orbital elements is no longer valid. In light of this, I proceed to use non-averaged tidal equations. These equations are obtained before the averaging of the satellite's radial distance $r$ over an orbital period, which would otherwise eliminate the true anomaly $f$ and lead to the appearance of the eccentricity functions $F_i(e)$ shown above. The non-averaged tidal equations for the evolution of $a$, $e$, $\omega$ and $\varepsilon$ are:

$$
\begin{aligned}
	\dot{a}= & \frac{2 K_{s}}{m_{s}} \frac {a^7} {r^8} \left\{ \frac {\omega_s}{n} \cos{\varepsilon_s} \sqrt{1-e^2} - \frac {3 e^2} {1-e^2} \sin^2{f} - \frac {a^2 (1-e^2)}{r^2} \right\} \\ \nonumber
	& + \frac{2 K_{p}}{m_{s}} \frac {a^7} {r^8} \left\{ \frac {\omega_p}{n} \cos{I} \sqrt{1-e^2} - \frac {3 e^2} {1-e^2} \sin^2{f} - \frac {a^2 (1-e^2)}{r^2} \right\}, \nonumber \\
	\dot{e}= & \frac{9 K_{s} a^6}{m_{s} r^7} \left\{\frac{\omega_s}{n}  \frac{X \sqrt{1-e^2}}{a} \cos{\varepsilon_s} - \frac {X a}{r^2} (1-e^2)  -\frac{e \sin^2{f}}{3 r} \right\} \nonumber \\
	& + \frac{9 K_{p} a^6}{m_{s} r^7} \left\{\frac{\omega_p}{n}  \frac{X \sqrt{1-e^2}}{a} \cos{I} - \frac {X a}{r^2} (1-e^2) -\frac{e \sin^2{f}}{3 r} \right\} , \nonumber \\
	\dot{\omega_s}= & -K_s \frac{n}{C_s} \frac{a^6}{r^6} \left\{\frac{\omega_s}{n} \{ \cos^2(o+f) + \cos{\varepsilon_s} \sin^2(o+f) \} - \frac{a^2 \sqrt{1-e^2}}{r^2} \cos{\varepsilon_s} \right\}, \nonumber \\
	\dot{\varepsilon}_{s}= & \frac{K_s}{C_s} \frac{a^6}{r^6} \left\{\frac{n}{\omega_s} \frac{a^2 \sqrt{1-e^2}}{r^2} \sin{\varepsilon_s} - \sin^2(o+f) \sin{\varepsilon_s} \cos{\varepsilon_s} \right\}, \nonumber \\ 
\end{aligned}
$$
\noindent where $f$ is the satellite's true anomaly, $o$ is the satellite's argument of the periapsis, and

$$X= \frac{2 \cos{f} + e (1+\cos^2{f})}{9 (1+e \cos{f})}, \nonumber$$

These equations are introduced in the integrator in a subroutine called at every $0.01$ day which defines the integration step length for the application of the tidal effect.

The initial conditions for the satellite are based on Figure 4 of \cite{nogueiraetal2011}. However, unlike the simulations that include only the satellite, we cannot explore a wide range of initial conditions due to the high computational cost. Therefore, for the semimajor axis, I choose the discrete value $0.07$ au (roughly $423 R_N$, which is near a great accumulation of captured satelites \citep{nogueiraetal2011}), and set the eccentricity such that $a_f = a(1 - e^2) = 0.003$ au. As shown in \cite{correia2009}, $a_f$ roughly corresponds to Triton's final semimajor axis and remains approximately constant when only the tidal equations are considered. In principle, a smaller $a_f$ would be required to stop Triton near its current distance from Neptune. However, due to the significant eccentricity variations induced by solar perturbations, test integrations showed that a larger $a_f$ is necessary; otherwise, most simulations would result in Triton's collision with Neptune.

The orbital inclination is set to $150^\circ$, measured with respect to Neptune's orbital plane. The mean anomaly is fixed at $180^\circ$, assuming the satellite is captured near the orbit's apoapsis. The longitude of the ascending node is varied from $0^\circ$ to $330^\circ$ in steps of $30^\circ$, and the argument of periapsis from $0^\circ$ to $350^\circ$ in steps of $10^\circ$. The Sun is assumed to follow a circular orbit around Neptune and is initially placed at the origin of longitudes, so that the satellite’s initial longitude of the ascending node is referenced to the Sun’s initial position.

I proceed with the integrations by including in the integration code a condition to stop the process whenever the satellite's apoapsis distance becomes smaller than $0.01$ au (approximately $60 R_N$)—cases in which the Sun's gravitational influence can be neglected. After this point, I continue the integration using only the secular, tidally averaged equations, as done in the previous subsection.

Due to the strong gravitational perturbations from the Sun, approximately $53$ \% of the orbits become eccentric enough to lead to Triton's collision with Neptune shortly after the beginning of the integration and before tidal forces can substantially reduce the eccentricity. The satellites that avoid early collision due to their highly eccentric orbits generally follow one of three evolutionary paths. They may:

1. Collide with Neptune after achieving a circular orbit and spiraling inward due to tidal decay before $4.5$ Gy,

2. Remain in an eccentric orbit at $4.5$ Gy, typically with a semimajor axis larger than Triton's current value, or

3. Reach a nearly circular orbit ($e < 0.001$) and decay to a semimajor axis roughly between $9 R_N$ and $18R_N$ at 4.5 Gy.

This last group can be considered Triton-like satellites and represents about $29$\% of all simulated cases. Although I did not perform extensive integrations with other initial semimajor axes for the satellite, some integrations for the satellite's semimajor axis equal to $0.2$ au (roughly $1200 R_N$) showed a fraction of satellites that ended near present Triton's semimajor axis of only $4.4$ \%. This suggests that Triton would most likely be captured with a not so large semimajor axis typically smaller than $10^3 R_N$. It is also possible that a larger value of $a_f$ could help prevent both early collisions (by increasing the initial periapsis) and late ones (since a larger $a_f$ would result in a greater semimajor axis once the orbit becomes circular). This possibility warrants further investigation, although it lies beyond the main scope of this paper.

\begin{figure}
\centering
\includegraphics[scale=.40, angle=-90]{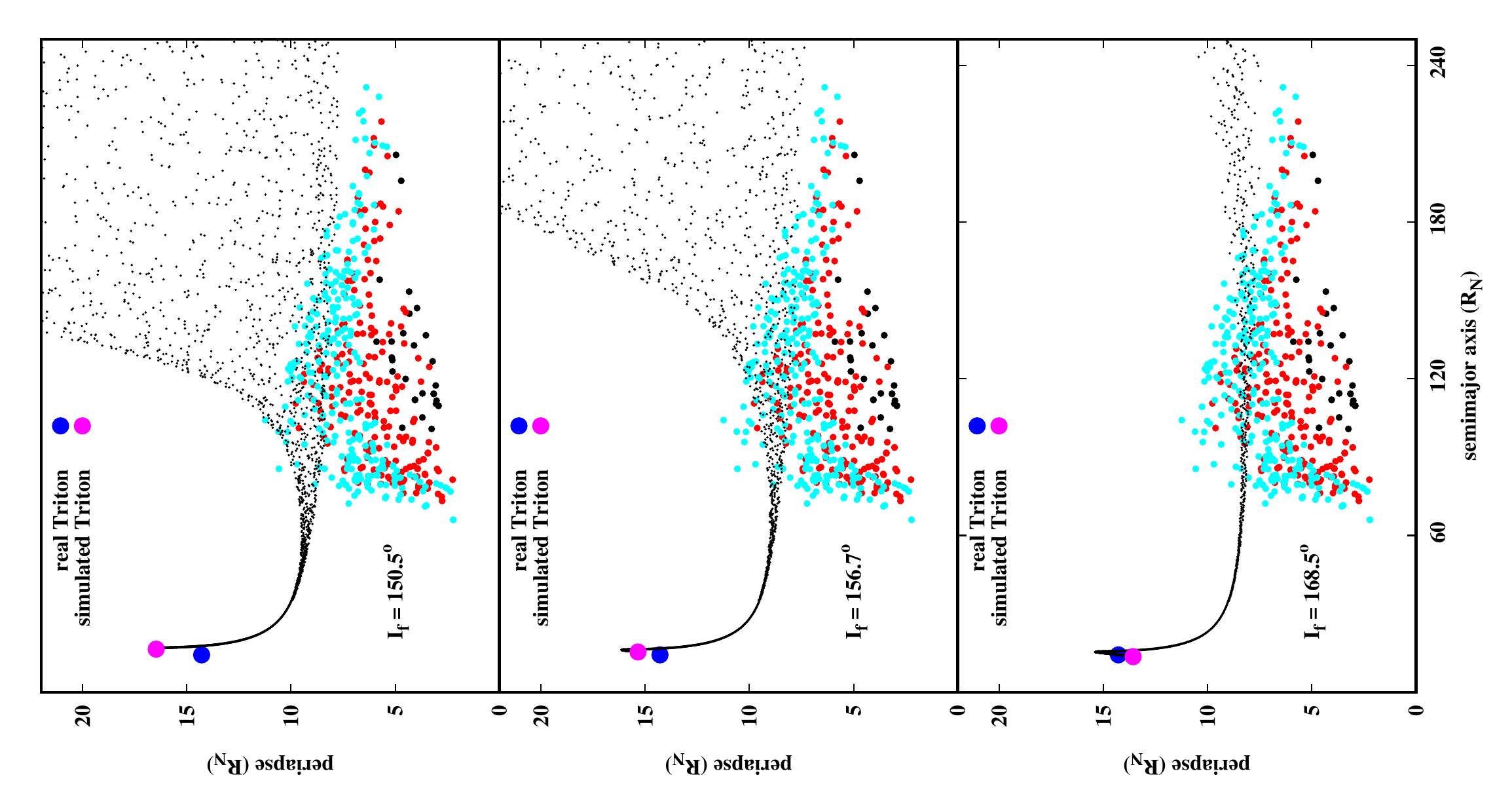}
	\caption{Evolution of the semimajor axis and periapsis of several satellites affected by tides (black small dots) and perturbed by the Sun, plotted over the map showing the locations where resonances between Neptune’s precession frequency and $s_8$ occur, as depicted in Fig. 6. The large blue circle represents Triton, and the large magenta circle represents the final position of the simulated satellite. The upper and lower panels illustrate the extreme cases with respect to the amplitude of periapsis fluctuations, which correspond to extreme values of the satellite’s final inclination $I_f$. The middle panel shows an intermediate case, roughly corresponding to Triton’s inclination.}
        \label{fig-aq-justsun}
\end{figure}

Returning to the main focus of this paper, I examine the evolution of the orbits that end in circular configurations near Triton’s current semimajor axis. All cases in which the satellite ended in a circular orbit near Triton ($10 R_N < a < 18 R_N$) showed the satellite's semimajor axis and periapsis passing through the secular resonance region, as depicted in Fig. \ref{fig-ressecs8-restr}. Three of these cases are illustrated in Fig. \ref{fig-aq-justsun}, which shows the evolution of the semimajor axis and periapsis for each. The top and bottom panels represent extreme cases with respect to the amplitude of periapsis fluctuations within the resonance region. It is noteworthy that the amplitude of periapsis variations is inversely proportional to the satellite’s inclination relative to Neptune’s equator. This relationship between the amplitude of eccentricity variations and inclination is linked to the Kozai resonance induced by the Sun. Inclinations near $90^\circ$ are associated with larger amplitude variations in both inclination and eccentricity, in contrast to inclinations close to $180^\circ$.

The middle panel shows a case in which the satellite ends with an inclination close to Triton’s. Although in all cases the satellite crosses the resonance region, when the inclination is higher (closer to $180^\circ$), the smaller amplitude variations in eccentricity (and hence in periapsis) result in a longer residence time of the semimajor axis–periapsis pair within the resonance region. In the case where the satellite’s inclination is close to Triton’s, the satellite fully crosses the resonance region for the semimajor axis spanning approximately from $80 R_N$ to $130 R_N$. Outside that range, the periapsis crosses the resonance region only partially, and only when it reaches lower values during its fluctuations.

\subsubsection{Satellite, Sun and the other major planets}
 
I now include the perturbations from the other three major planets. Based on the results of the previous section, I select the initial orbital elements of the satellite that yield a final semimajor axis close to Triton’s and an eccentricity near zero ($e < 10^{-3}$). Owing to the high computational cost, only a subset of the eligible cases from the previous section was chosen for this new set of simulations.

The planets are added according to coordinates referred to Neptune's center based on data from JPL Horizons\footnote{\url{https://ssd.jpl.nasa.gov/horizons/app.html\#}}. To ensure compatibility with the simulations of the previous section, I set Neptune’s initial eccentricity to zero and choose the reference frame to coincide with Neptune’s orbital plane. The coordinates and velocities of the other planets are selected to preserve the same relative configuration with respect to Neptune, and the satellite’s coordinates and velocities are chosen to maintain the same relative configuration with Neptune as in the cases selected from the previous section. Note that, in the present case, in addition to Neptune’s initial circular orbit, its initial obliquity is also set to zero, unlike in the first exploratory integrations described in Section 3.2, so that consistency with the integrations without the major planets can be maintained.

Although all cases simulated in the previous section assumed a satellite tidal response time of $808\,\mathrm{s}$, I also test other values for this parameter, ranging from $1000\,\mathrm{s}$ to $4000\,\mathrm{s}$. This choice is justified because, when the satellite approaches Neptune too closely at periapsis prior to orbital circularization, it must undergo significant tidal heating, consistent with a semi-molten state and therefore implying a larger tidal response time.

\begin{figure}
\centering
\includegraphics[scale=.38, angle=-90]{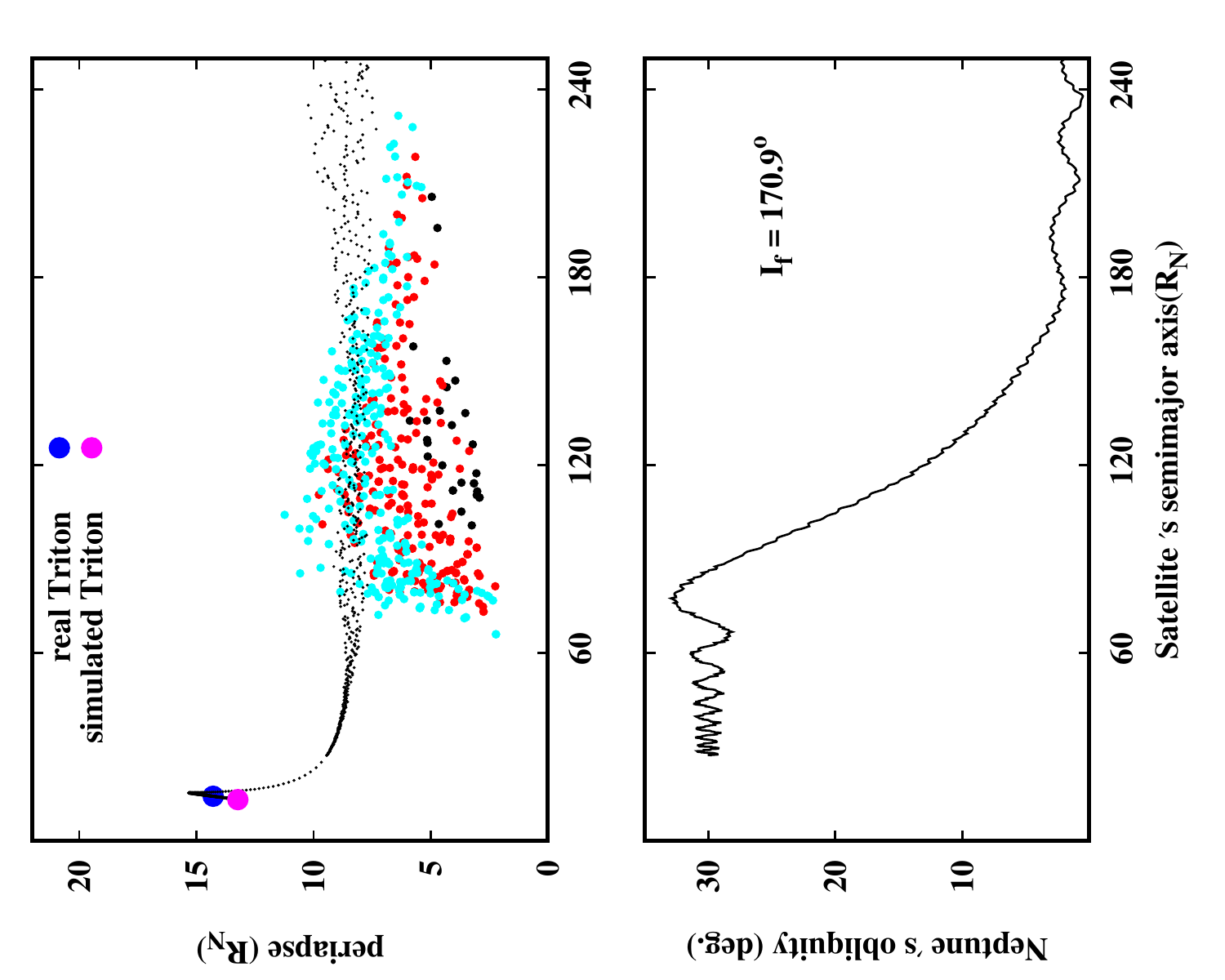}
	\caption{Upper panel: Evolution of the satellite’s semimajor axis and periapsis under tidal evolution, depicted by small black circles. The other circles and their colors are explained in the labels of Fig.~\ref{fig-ressecs8-restr}. The pair ($a$,$q$) crosses the resonance-inducing region well within its boundaries. For semimajor axes of the satellite around $25 R_N$ and smaller, the integration is performed considering only the satellite’s tidal evolution, neglecting the Sun and the other planets; this can be observed in the lower density of points in the $a–q$ evolution. The influence of the satellite’s evolution shown in the upper panel on Neptune’s obliquity can be seen in the lower panel, where Neptune’s obliquity is plotted as a function of the satellite’s semimajor axis. Neptune’s final obliquity slightly exceeds the observed present value. The satellite that produced this result had initially its longitude of the ascending node equal to $90^\circ$, argument of the perispsis equal to $190^\circ$ and $\Delta t_s = 2500\,\mathrm{s}$.} 
	\label{asat-oblnet-good}
\end{figure}

Figure~\ref{asat-oblnet-good} shows one of the most interesting cases obtained from these all–giant-planet simulations. The upper panel, as in previous figures, illustrates the evolution of the satellite’s semimajor axis and periapsis, highlighting its passage through the resonance-inducing region. The lower panel shows the variation of Neptune’s obliquity as a function of the satellite’s semimajor axis. The final obliquity of Neptune is very close to the observed value. As far as time is concerned, Neptune's obliquity in the case of Fig. \ref{asat-oblnet-good} began to increase after about $5 \times 10^7$ years, and the total increase took about $2 \times 10^7$ years to complete.

Although not enough simulations were performed to obtain robust statistics, several important observations can be made. First, when the other three giant planets are included, not all cases that previously resulted in a surviving satellite on a nearly circular orbit reproduced the same outcome. In fact, some of the favorable cases from the previous section became cases in which the satellite collided with Neptune due to tidal evolution after reaching a circular orbit.

Although obliquities larger than that shown in Fig.~\ref{fig-aq-justsun} were obtained (with a maximum of $50.8^{\circ}$), many cases exhibited only a modest increase in Neptune’s obliquity. Cases in which Neptune’s obliquity exceeded $20^{\circ}$ accounted for $28$\% of all simulations that resulted in the satellite in a final circular orbit near Triton. Cases that induced the largest obliquities for Neptune are associanted with satellites initially with the longitude of the node at or near $90^\circ$ or $270^\circ$. 

Another noteworthy result is that larger obliquities of Neptune were obtained when the satellite’s final inclination was close to $170^{\circ}$ or higher. This outcome is expected, since larger satellite inclinations are associated with a smaller amplitude of periapsis periodic fluctuations, allowing the satellite to traverse a wider portion of the resonance-inducing region.

\section{Conclusions}

I have shown that if Triton was originally captured by Neptune through the binary planetesimal rupture model, its subsequent tidal evolution would, with high probability, cause its orbit to cross a region that—together with the perturbations from the Sun and the other major planets—induces an increase in Neptune’s obliquity through the action of a spin–$s_8$ resonance. The passage of the satellite through this resonance-inducing region does not affect Neptune’s obliquity uniformly. Higher obliquities are obtained for satellites whose inclination with respect to Neptune’s equator approaches $180^{\circ}$, which naturally implies a smaller amplitude of eccentricity periodic fluctuations due to solar perturbations. This, in turn, leads to an evolution of the semimajor axis–eccentricity pair that remains largely within the resonance-inducing region.

Although only a limited number of simulations were performed with the complete model (including tides, the Sun, and the other three major planets), a tentative estimate indicates that approximately $28$\% of all cases that resulted in the satellite in a final circular orbit near Triton produced an increase in Neptune’s obliquity greater than $20^{\circ}$. The largest obliquity obtained reached $50.8^{\circ}$. It is important to point out here that \cite{saillenfest-lari2021} (also \cite{saillenfestetal2023}) derive an inequality involving the mass of a satellite orbiting a planet that could trigger a spin–orbit resonance with a given nodal harmonic. According to this criterion, Triton would be too low in mass to cause any significant effect on Neptune’s obliquity. The missing factor is that the authors’ criterion does not involve the satellite’s eccentricity, and here I show that the eccentricity plays a fundamental role in triggering the spin–orbit resonance.

Although several other mechanisms have been proposed to tilt Neptune’s spin axis, the scenario developed in this paper has the advantage of not invoking any ad hoc assumptions, being based solely on its actual moon Triton and its likely tidal evolution. In this sense, the scenario presented here also reinforces the hypothesis of Triton’s capture through the binary planetesimal rupture mechanism \citep{Agnor-Hamilton2006} which results in an initially highly eccentric orbit for Triton.

There is still ongoing debate about whether the formation of the ice giants involved collisions between planetary embryos or relied solely on pebble accretion. The latter model naturally implies final planetary obliquities close to zero. The fact that Neptune’s obliquity can be naturally explained by the evolution of the Neptune–Triton system alone further reinforces a formation scenario for Neptune that does not require any major primordial collision with a massive embryo. Also, since Uranus has a rotation period quite similar to that of Neptune, it is not likely that, after the onset of the putative embryos' collisions with the ice giants, they would end up with similar rotation periods. The fact that I show here that Neptune's obliquity can be attained without any recourse to catastrophic events also suggests that some non-catastrophic process may have tilted Uranus’s spin axis to its present inclination. Moreover, the results presented here also support the idea that exoplanets, in general, are not expected to have near-zero obliquities.

Another noteworthy point concerns the time at which Neptune's obliquity begins to be excited. For the case depicted in Fig. \ref{asat-oblnet-good}, it took $5\times10^7$ years for Neptune's obliquity to start being excited. The simulations performed with all the major planets that could induce the spin–$s_8$ resonance were started with the satellite's semimajor axis at about $400 R_N$, and the excitation of Neptune's obliquity occurred in less than $0.1$ Gy. However, other integrations started with the satellite's semimajor axis between $1000$ and $2000$ au showed the beginning of the satellite's circularization after several hundred million years to about one Gyr. Thus, depending on the actual orbit of Triton just after its capture, we would expect that Neptune's obliquity excitation could begin at any time from a few tens of millions of years to about one Gyr. This aspect of the scenario presented here differs fundamentally from other scenarios in which the time window for the excitation of Neptune's obliquity is strongly restricted.

As a final comment, \cite{hammond-collings-2024} recently argued that internal feedbacks in Triton between tidal heating and ice-shell melting significantly reduce the rate of orbital evolution, thereby extending the time Triton remains on a highly eccentric orbit. This suggests that Triton would spend a longer time in the resonance-inducing region, naturally allowing Neptune to experience significant variations in its obliquity over an extended period. This point deserves further investigation in future extensions of the survey undertaken in this paper.

\appendix

\section{Introduction of Euler's Equations in a Numerical Integration of the Equations of Motion}

The variation of Neptune's obliquity is addressed by incorporating Euler's equations for a rigid body. The approach involves integrating Euler’s equations alongside the Newtonian equations of motion for an N-body system. This will be done utilizing the classical integrator RADAU \citep{everhart1985}, since the method here developed does not support an association with a sympletic integrator. Euler’s equations are applied to the central body, while all objects gravitationally interact with one another and also influence Neptune’s spin axis.

To address the specific problem of the variation of Neptune’s spin axis under the influence of other objects (Triton, the Sun, and the other planets), a general framework is developed. This framework applies to any central body treated as a rigid body influenced by other objects modeled as point masses. The central body need not be the most massive in the system, but its center of mass must be the origin of the reference frame.

Euler’s equations for a rigid body can be written as:

\begin{eqnarray}
        & A \dot \omega_x - (B-C) \omega_y \omega_z & = 3 G m (C-B) \frac {y z} { r^5 } \nonumber\\
        & B \dot \omega_y - (C-A) \omega_x \omega_z & = 3 G m (A-C) \frac {x z} { r^5 } \nonumber\\
        & C \dot \omega_z - (A-B) \omega_x \omega_y & = 3 G m (A-B) \frac {x y} { r^5 } \nonumber\\\nonumber
\end{eqnarray}\nonumber

\noindent where $A$, $B$ and $C$ are the principal moments of inertia of the central body, $\vec \omega$ is its spin vector, $G$ is the gravitational constant, $x$, $y$ and $z$ are the rectangular coordinates of the perturbing object, $r$ is its the radial distance and $m$ is its mass.

To further simplify the equations, the central body is assumed to be axially symmetric, which is a good approximation for large bodies such as planets. Under this assumption, Euler’s equations reduce to:

\begin{align}
         \dot \omega_x & = 3 G \frac {C-A} {A} \sum_{i=1}^N { m_i \frac {y_i z_i} {r_i^5} = S_x} \nonumber\\
         \dot \omega_y & = -3 G \frac {C-A} {A} \sum_{i=1}^N { m_i \frac {x_i z_i} {r_i^5} = S_y} \label{eq1} \tag{1} \\
\end{align}

\noindent where I add the perturbations for N bodies.

Although Euler’s equations are relatively simple compared to more general formulations (such as Andoyer equations), they are valid only in a reference frame whose axes align with the principal axes of inertia of the body. Consequently, after the first integration step, the rotation axis of the central body shifts. In the gyroscopic approximation adopted here, the principal axis of inertia follows the rotation axis, rendering Euler’s equations invalid for the next step.

To address this issue, a reference frame transformation is performed at each integration step, ensuring that the equations remain valid throughout the simulation. At every step, the vector $\vec \omega$ is shifted from its vertical orientation by adding the differential components $d \omega_x$ and $d \omega_y$, given by:

 \begin{eqnarray}
        & d \omega_x & = \int_0^t { S_x dt } \nonumber \\
        & d \omega_y & = \int_0^t { S_y dt } \nonumber \\ \nonumber
\end{eqnarray}

\noindent where $S_x$ and $S_y$ are functions of the coordinates of the perturbing bodies. In principle, $d \omega_x$ and $d \omega_y$ can be computed through a simple first-order integration of $S_x$ and $S_y$, assuming the coordinates of the objects remain constant and equal to their values prior to the current integration step of the Newtonian equations of motion. However, this approach was found to introduce systematic errors in the evolution of the central body's spin axis. To overcome this, I implemented a scheme in which the Newtonian equations are integrated simultaneously with Euler’s equations at every step, as described below.

The rotation matrix used to update the reference frame at each step is constructed from a sequence of rotations: first, a rotation around the Z-axis by an angle $\Omega$, followed by a rotation around the X-axis by an angle $\gamma$, and finally a rotation around the Z-axis by an angle $-\Omega$. The angles $\Omega$ and $\gamma$ are related to $d \omega_x$ and $d \omega_y$ by:

\begin{eqnarray}
        & \cos{\Omega} & = \frac {d \omega_x} {d \omega_{xy}} \nonumber \\
        & \sin{\Omega} & = \frac {d \omega_y} {d \omega_{xy}} \nonumber \\
        & \cos{\gamma} & = \frac {d \omega_{xy}} {\omega} \nonumber \\
        & d \omega_{xy} & = \sqrt{ d \omega_x^2 + d \omega_y^2} \nonumber \\ \nonumber
\end{eqnarray}

\noindent and the rotation matrix $M$ thus writes:

$$
M =
\begin{vmatrix}

        1- \frac {d \omega_x^2} {2 \omega^2} & - \frac{d \omega_y^2} {2 \omega^2} & - \frac{d \omega_x}{w} \\
        - \frac{d \omega_y^2} {2 \omega^2}  & 1- \frac {d \omega_y^2} {2 \omega^2}  & - \frac{d \omega_y}{w}  \\
        \frac{d \omega_x}{w} & \frac{d \omega_y}{w}  & 1- \frac {d \omega_{xy}^2} {2 \omega^2} 

\end{vmatrix}
$$

I now proceed to incorporate the reference frame rotation into the numerical integration of the equations of motion. The process begins with a reference frame in a given orientation, and perturbing objects at known positions and velocities relative to this frame. After one integration step, the updated positions and velocities of the objects must be expressed with respect to the rotated reference frame. To implement this, I define the reference frame rotation by the equation $\vec r_t = M \vec r$, where $\vec r$ is the radius vector of a perturbing body before the rotation, $r_t$ is the radius vector after the rotation, and $M$ is the rotation matrix. I then take the first and second derivatives of $\vec r_t$:

\begin{align}
        \dot \vec r_t & = \dot M \, \vec r + M \, \dot \vec r = \vec v_t  \label{eq2} \tag{2} \\
         \ddot \vec r_t & = \ddot M \, \vec r + 2 \dot M \, \vec v + M \, \ddot \vec r \label{eq3} \tag{3} \\
\end{align}

\noindent where $\vec v = \dot \vec r$ is the osculating velocity. To incorporate the reference frame rotation into a classical integrator, we must replace $\ddot \vec r$ with $\ddot \vec r_t $, whose expression includes the Newtonian acceleration $\ddot \vec r$, the osculating velocity $\vec v$, the radius vector $\vec r$, and the first and second derivatives of the rotation matrix $M$. In order for a second-order integrator to function correctly, it is essential to express $\ddot \vec r_t $ as a function of $\vec v_t$ (and not $\vec v$). Therefore, we must instead use the following expression for $\ddot \vec r_t$:

\begin{equation}
        \ddot \vec r_t  = \ddot M \, \vec r + 2 \dot M \, M^{-1} (\vec v_t - \dot M \vec r) + M \, \ddot \vec r \label{eq4} \tag{4} 
\end{equation}

\noindent where I replaced $\vec v$ in Eq. \ref{eq3} with $\vec v_t$ using Eq. \ref{eq2}\footnote{The vector $\vec r$ can be interpreted as $\vec r_t$ in Eq. \ref{eq3}, since it always represents the radius vector in the instantaneous reference frame. Thus, its derivative can be viewed either as $\vec v$ or $\vec v_t$, depending on whether we consider only the variation due to Newtonian perturbations or include the additional variation due to the reference frame rotation.}. To employ Eq. \ref{eq4} in the integration, we need to compute $\dot M$, $\ddot M$ and $M^{-1}$. These are functions of $d \omega_x$, $d \omega_y$, $d \omega_{xy}$, $S_x$, $S_y$, $\dot S_x$ and $\dot S_y$. Since $d \omega_x$, $d \omega_y$ and $d \omega_{xy}$ are differential elements, they can be neglected in comparison to $S_x$, $S_y$, $\dot S_x$ and $\dot S_y$. Therefore, we can take $M = M^{-1} = I$ and:

$$
\dot M =
\begin{vmatrix}

        0  & 0 & -\frac {S_x} {\omega} \\
        0  & 0 & -\frac {S_y} {\omega} \\
        \frac {S_x} {\omega} & \frac {S_y} {\omega}  & 0 

\end{vmatrix}
$$

$$
\ddot M =  
\begin{vmatrix}

        -\frac {S_x^2} {\omega^2}  & -\frac {S_x S_y} {\omega^2} & -\frac {\dot S_x} {\omega} \\
        -\frac {S_x S_y} {\omega^2} & -\frac {S_y^2} {\omega^2}  & -\frac {\dot S_y} {\omega} \\
        \frac {\dot S_x} {\omega} & \frac {\dot S_y} {\omega}  & -\frac {S_{xy}^2} {\omega^2}  

\end{vmatrix}
$$

\noindent where $S_x$, $S_y$ are given by Eqs. \ref{eq1}, $S_{xy}^2 = S_x^2+S_y^2$, and:

\begin{eqnarray}
        & \dot S_x & = 3 G \frac {C-A} {A} \sum_{i=1}^N { \frac {m_i} {r_i^5} (z_i {v_y}_i + y_i {v_z}_i -5 P_i \frac {y_i z_i} {r_i^2})} \nonumber \\
        & \dot S_y & = 3 G \frac {C-A} {A} \sum_{i=1}^N { \frac {m_i} {r_i^5} (z_i {v_x}_i + x_i {v_z}_i -5 P_i \frac {x_i z_i} {r_i^2})} \nonumber \\
        & P_i & = x_i {v_x}_i + y_i {v_y}_i + z_i {v_z}_i \nonumber  \\ \nonumber 
\end{eqnarray}

The final expression for $\ddot r_t$ becomes:

\begin{equation}
        \ddot \vec r_t  = (\ddot M - 2 \dot M^2) \vec r + 2 \dot M \vec v_t + \ddot \vec r \nonumber 
\end{equation}

The term $(\ddot M - 2 \dot M^2) \vec r + 2 \dot M \vec v_t$ stands for the reference frame correction and must be added to the Newtonian acceleration $\ddot \vec r$ in the Force subroutine of the numerical integrator. Their components $f_x$, $f_y$, $f_z$ are:

\begin{eqnarray}
        & f_x & = T_x^2 x + T_x S_y y - \dot T_x z -2 T_x v_z \nonumber \\
        & f_y & = T_x T_y x + T_y^2 y -\dot T_y z - 2 T_y v_z \nonumber \\
        & f_z & = \dot T_x x +\dot T_y y + T_{xy}^2 z + 2 (T_x v_x + T_y v_y) \nonumber \\ \nonumber 
\end{eqnarray}

\noindent where $T_{x,y,xy} = S_{x,y,xy}/\omega$ and $\dot T_{x,y,xy} = \dot S_{x,y,xy}/\omega$

The construction of this integrator ensures that its output is always given with respect to the planet’s equatorial frame. However, the output can be transformed to describe orbits in any other reference frame. In particular, we can track the variation of the planet’s equator relative to the initial reference frame—defined as the planet’s equator at $t=0$ by introducing a neutral, massless satellite in a circular orbit initially lying in the planet's equatorial plane. This satellite is assumed to be influenced only by a spherically symmetric component of the planet’s gravitational field, while the "perturbations" it experiences are purely those introduced by the reference frame rotation equations. As a result, the satellite's orbit remains fixed in the initial equatorial plane, allowing us to define the evolving orientation of the equator relative to the initial frame at any point in time.

For all simulations presented in the following sections, the reference frame rotation components are implemented within the FORCE subroutine of the 15th-order RADAU integrator \citep{everhart1985}, using an accuracy parameter of $12$.

\section{Validation tests}

\begin{figure}
\centering
\includegraphics[scale=.30, angle=-90]{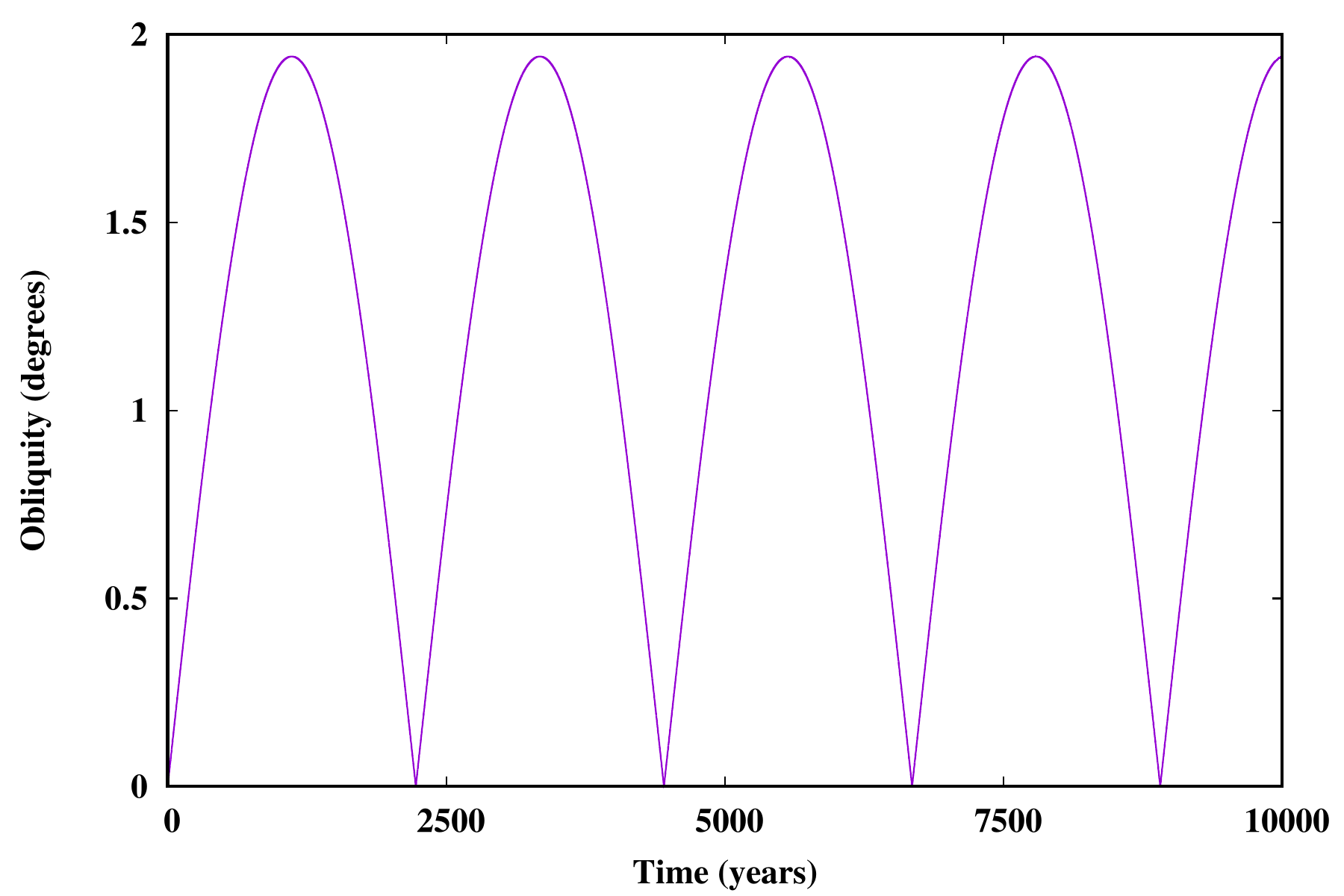}
        \caption{Variation of the obliquity of a planet affected gravitationally by a satellite. The obliquity is measured from the reference frame coincinding with the planet's equator at time $t=0$. Other details in the main text.}
        \label{fig-valid1}
\end{figure}

I conduct two experiments to test the method described in the previous section. The first employs a simplified model that can be analytically validated. I consider a planet whose shape is characterized by the parameter $J_2$, and a satellite in orbit around it. The satellite's orbit is assumed to be circular and lies in a plane inclined relative to the planet’s equator. The parameters for the planet are: mass $M=5 \times 10^{-5} m_{\odot}$, flattening parameter $J_2=3.4 \times 10^{-3}$, equatorial radius $R=1.6e-4$ au, moment of inertia factor $0.23$, and spin frequency $3442$ yr$^{-1}$. The satellite has a mass of $10^{-8} M_{\odot}$, a semimajor axis of $0.003$ au, and an orbital inclination of $45^{\circ}$ relative to the planet’s equator. I integrate this system over $10^5$ years and numerically compute both the amplitude and the period of the planet’s obliquity variation from the simulation output. Figure \ref{fig-valid1} shows this variation, from which I determine an amplitude of $1.9412^{\circ}$ and a period of $2226.4$ years. This simple problem also admits an analytical solution. The amplitude of the obliquity variation ($L$) \footnote{Both the numerically and analytically calculated amplitudes are considered here with respect to the planet's initial equator, which corresponds to twice the amplitude relative to the invariable plane.} and its period ($p$) are given by the expressions:

\begin{eqnarray}
        & L & =  2 \arcsin{ \frac {\beta \sin{I}} {M_t} } \nonumber \\
        & p & =  2 \pi / (\alpha \frac {M_t} {\gamma \beta} \cos{I}) \nonumber \\
        & M_t & = \sqrt{\gamma^2 + \beta^2 + 2 \gamma \beta \cos{I} } \nonumber \\
        & \alpha & = 1.5 \frac {G} {a^3} (C-A) m \nonumber \\ \nonumber 
\end{eqnarray}

Here, $\gamma$ denotes the planet’s rotational angular momentum, $\beta$ is the satellite’s orbital angular momentum, $I$ the orbital inclination relative to the planet’s equator, $a$ the orbital semimajor axis, $C$ and $A$ are the planet’s principal moments of inertia, and $m$ is the satellite’s mass. The expression for $L$ is derived from the geometry of the angular momentum vectors, while the expression for $p$ is based on the value of $\Omega_0$ as given in \cite{bouelaskar2006}. These expressions yield an amplitude $L=1.9416^{\circ}$ and a period of $p=2226.9$ years. The discrepancies between the numerical and analytical results appear only in the fifth decimal place, consistent with the accuracy observed in most of the validation runs I performed.

\begin{figure}
\centering
\includegraphics[scale=.30, angle=-90]{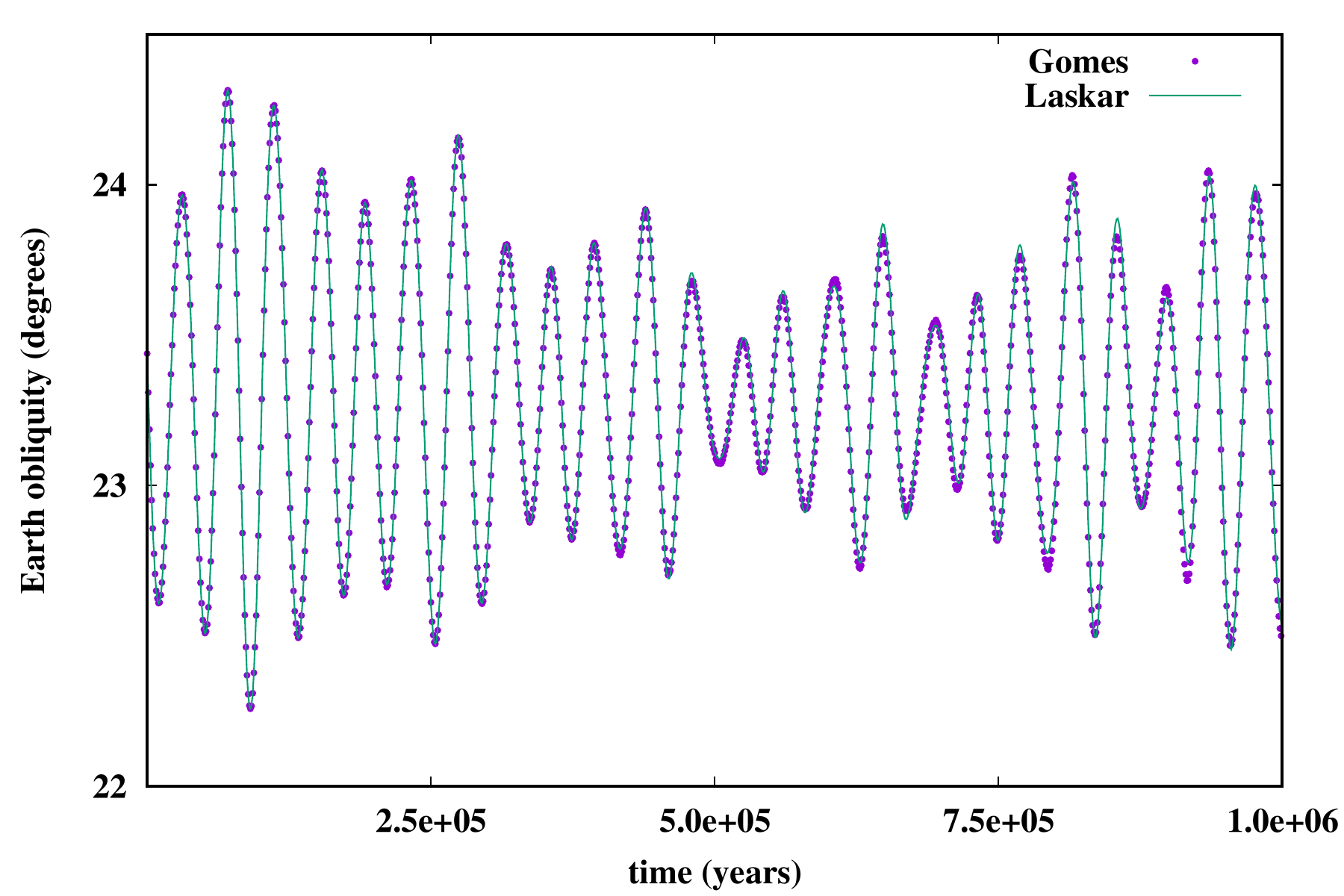}
        \caption{Variation of the obliquity of the Earth affected gravitationally by the Moon, Sun and the other seven planets compared with the obliquity taken from http://vo.imcce.fr/insola/earth/online/earth/La2004/INSOLP.LA2004.BTL.ASC, data generated according to \cite{laskaretal2004}}
        \label{fig-valid2}
\end{figure}

The second validation test involves a more complete system, with the Earth as the central rigid body and the Moon, the Sun, and the other seven planets modeled as point masses. The integration begins with the geocentric coordinates of all bodies at epoch J2000, obtained from JPL Horizons. Physical parameters, where applicable, are taken from \cite{laskaretal2004}, whose results are used for comparison. Figure \ref{fig-valid2} shows the comparison of the obliquity of the ecliptic over $10^6$ years. The agreement is satisfactory, especially considering that \cite{laskaretal2004} includes dissipative effects not accounted for in this work. The maximum relative difference between the two methods over the first $10^6$ years is $4 \times 10^{-3}$.

\vskip 200pt

\bibliographystyle{aasjournal}

\bibliography{bib}

\end{document}